# Switching probability investigation of electric field-induced precessional magnetization switching

**Hongguang Cheng and Ning Deng**

Institute of Microelectronics, Tsinghua University, Beijing 100084, P. R. China

E-mail: chenghg7932@gmail.com

**Abstract**

We report theoretical investigation of the switching probability of electric field-induced precessional magnetization switching by solving the Fokker-Planck equation numerically with finite difference method. The switching probability is determined by the net magnetic field induced by the deviation of precession angle from its equilibrium position after precession process. The error rate has the lowest value under an appropriate applied external field for the voltage pulse duration $\tau_{\text{pulse}}$ a little longer than the half precession period. The calculated results show that ultra-low error rate down to the order of $10^{-12}$ can be achieved for thermal stability factor $\Delta = 50$ and low damping factor material should be used for free layer to improve the switching probability. For parallel (anti-parallel) magnetization to anti-parallel (parallel) magnetization switching process, the spin transfer torque tends to decrease (increase) the error rate when the $\tau_{\text{pulse}}$ is shorter than the half precession period, and increase (decrease) the error rate when $\tau_{\text{pulse}}$ is longer than the half-period. These results exhibit potential of electric field-induced precessional magnetization switching for ultra-low power, high speed magnetic random access memory (MRAM) application.

## 1. Introduction

Electric field-induced precessional magnetization switching utilizing the interfacial voltage-controlled magnetic anisotropy [1-12] to modify the free layer perpendicular anisotropy field has been demonstrated recently [13-16]. A bistable magnetization switching with sub-nanosecond switching time is realized by applying a unipolar voltage pulse in FeCo/MgO/Fe magnetic tunnel junction [13, 15]. Its realization in CoFeB/MgO materials system [14, 16] is of most technological importance for the capability of high density integration with conventional semiconductor industry and large tunnel magnetoresistance ratio [17, 18]. Magnetization reversal induced by electric field only consumes charging and discharging energy, it can reduce writing power consumption by more than two orders of magnitude compared with the spin transfer torque (STT) induced switching. Therefore, electric field-induced precessional magnetization switching is a promising candidate for ultra-low power and high speed magnetic random access memory (MRAM) applications.

Due to the thermal fluctuations of magnetization direction, switching error may take place in the switching process. The switching probability is defined by the number of error-free switching events divided by the number of total switching events. It characters the reliability of device write operation. Error rate is one minus switching probability. Ultra-low error rate should be achieved for high-reliability applications. In STT induced switching, the error rate can be unlimited close to zero by increasing the write duration time. While in electric field-induced precessional magnetization switching, the error rate is a limited value. The order of error rate can be achieved and the factors influence the switching probability need to be made clear. In previous studies [13-16] the switching probability is obtained qualitatively by repeating macro-spin model simulation or micro-magnetic simulation hundreds of times. This could not be used to investigate the switching events with ultra-low error rate quantitatively. Although the error rate can be obtained by experimental approach [19, 20], theoretical investigation of error rate quantitatively is necessary to design the electric field-induced precessional magnetization switching device.

In this paper, we report quantitative calculations of the switching probability of electric field-induced precessional magnetization switching with in-plane easy axis [13, 15] by solving the Fokker-Planck equation [21] numerically with finite difference method based on macro-spin model. We investigated the mechanism that switching errors take place and discussed the influence of thermal stability factor, damping factor of free layer and spin transfer torque effect on the switching probability.

## 2. Model and methods

Figure 1 demonstrates the schematic geometry of the electric field-induced precessional magnetization switching. $\theta$ and $\varphi$ are the direction angles describing the orientation of the macro-spin moment **M** of free layer in spherical coordinates. The in-plane coercive field $\mathbf{H}_c$ is parallel to the $x$ axis. The external magnetic field $\mathbf{H}_{ext}$ is applied along the positive direction of $z$ axis. The calculated trajectories by macro-spin model simulation based on zero temperature Landau-Lifshitz-Gilbert (LLG) equation are shown in figure 1(a)-(b). $\mathbf{M}_{initial}$ and $\mathbf{M}_{final}$ denote the initial and final magnetization state before and after voltage pulse duration $\tau_{pulse}$. The blue line represents the precession process during $\tau_{pulse}$, and the red line represents the relaxation process after $\tau_{pulse}$.

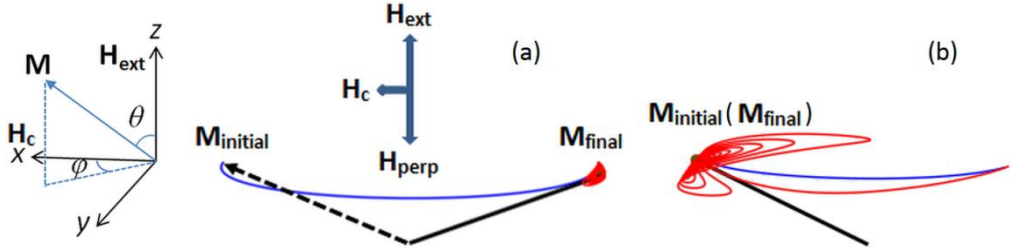

**Figure 1.** Schematic geometry of the electric field-induced precessional magnetization switching. $\theta$ and $\varphi$ are the direction angles describing the orientation of the macro-spin moment **M** of free layer in spherical coordinates. The in-plane coercive field $\mathbf{H}_c$ is parallel to the $x$ axis. The external field $\mathbf{H}_{ext}$ is applied along the positive direction of $z$ axis. (a) and (b) are the calculated trajectories based on zero temperature LLG equation. $\mathbf{M}_{initial}$ and $\mathbf{M}_{final}$ denote the initial and final magnetization state before and after voltage pulse duration $\tau_{pulse}$. Blue line represents the precession process during $\tau_{pulse}$, and red line represents the relaxation process after $\tau_{pulse}$.

Before the voltage pulse is applied, the initial magnetization state $\mathbf{M}_{initial}$ is along the direction of effect field $\mathbf{H}_{eff}=\mathbf{H}_c+\mathbf{H}_{perp}+\mathbf{H}_{ext}$, where $\mathbf{H}_{perp}$ is the out-of-plane anisotropy field. When apply a voltage pulse, a decrease of $H_{perp}$ is induced by the electrical field modulation

[13], this will produce a net magnetic field *H* in the *z* positive direction:

$$H = H_{\text{ext}} - H'_{\text{perp}} \cos\theta, \tag{1}$$

where $\theta$ is the precession angle, $H'_{\text{perp}}$ is the magnitude of out-of-plane anisotropy field when the voltage pulse is applied. Magnetization precession will take place under this induced net magnetic field.

Due to the non-zero in-plane coercive field **H**$_c$, the precession axis tilts from the *z* axis. During the precession process, the **H**$_c$ changes from $H_c$ to $-H_c$, the direction of precession axis will move, the precession angle $\theta$ will decrease in the first quarter of precession period and then increase in the later quarter of period. In this process, the precession angle $\theta$ also decreases gradually due to the damping effect. This will produce a net magnetic field *H'* in the *z* negative direction after $\tau_{\text{pulse}}$:

$$H' = H_{\text{perp}} \cos(\theta_0 - \Delta\theta) - H_{\text{ext}} \approx H_{\text{perp}}[\cos(\theta_0 - \Delta\theta) - \cos\theta_0], \tag{2}$$

where $\theta_0$ is the precession angle at initial equilibrium position and $\Delta\theta$ is the deviation of precession angle from its equilibrium position after $\tau_{\text{pulse}}$. The net magnetic field *H'* will lead to a reversed magnetization precession. It is an increasing function with respect to $\Delta\theta$. When turn off the voltage pulse after the duration time nearby half precession period, if the $\Delta\theta$ is not large, under the effect of **H'** and in-plane coercive field **H**$_c$, the magnetization will swing around the equilibrium position and the final magnetization state will be stabilized to the reversed orientation with respect to the initial state during the relaxation process, as shown in figure 1(a). Therefore, a coherent magnetization switching can be realized by a unipolar voltage pulse. If the $\Delta\theta$ is large enough, the reversed precession will make the final magnetization state back to its initial orientation during the relaxation process, as shown in figure 2(b), so a switching error takes place.

To calculate the magnetization switching probability, we need to know the magnetization probability density distribution $P(\theta,\varphi)$ after the switching process. The switching process consists of the precession process during $\tau_{\text{pulse}}$ and the relaxation process after $\tau_{\text{pulse}}$. The time-dependent $P(\theta,\varphi,t)$ follows the Fokker-Planck equation [21]:

$$\frac{\partial P}{\partial t} = \frac{1}{\sin\theta}\frac{\partial}{\partial\theta}\left\{\sin\theta\left[\left(h\frac{\partial V}{\partial\theta} - g\frac{1}{\sin\theta}\frac{\partial V}{\partial\varphi}\right)P + k\frac{\partial P}{\partial\theta}\right]\right\} + \frac{1}{\sin\theta}\frac{\partial}{\partial\varphi}\left\{\left(g\frac{\partial V}{\partial\theta} + h\frac{1}{\sin\theta}\frac{\partial V}{\partial\varphi}\right)P + k\frac{1}{\sin\theta}\frac{\partial P}{\partial\varphi}\right\}, \tag{3}$$

where $h = \frac{\alpha\gamma}{M_s + M_s\alpha^2}$, $g = -\frac{\gamma}{M_s + M_s\alpha^2}$, $k = k_B T h/\nu$, $\alpha$ is the damping factor, $\gamma$ is the gyromagnetic ratio, $k_B$ is the Boltzmann constant, $T$ is the temperature, $M_s$ is the saturation magnetization, $v$ is the volume of free layer and $V$ is the effective potential energy under effect field $\mathbf{H}_{\text{eff}}$. $V$ is given by: $V = -\mathbf{M} \cdot \mathbf{H}_{\text{eff}} = (-H_{\text{perp}} M_s \cos^2\theta/2 - H_{\text{ext}} M_s \cos\theta - H_c M_s \sin^2\theta \cos^2\varphi/2)v$. The initial probability density distribution $P(\theta,\varphi,0)$ is in its thermal equilibrium state, i.e., for $-\pi/2 \leq \varphi \leq \pi/2$, $P(\theta,\varphi,0)$ is given by the Boltzmann distribution $P_0 \exp(-V/k_B T)$, and for $\pi/2 < \varphi < 3\pi/2$, $P(\theta,\varphi,0) = 0$, where $P_0$ is the normalization constant: $P_0 = 1/\int_0^\pi d\theta \int_{-\pi/2}^{\pi/2} M_s^2 \sin\theta \exp(-V/k_B T) d\varphi$. Then the problem addressed is solving the partial differential equation (3). Once the probability density distribution $P(\theta,\varphi)$ after the switching process is obtained, the switching probability $P_{\text{switch}}$ can be obtained by integral:

$$P_{\text{switch}} = \int_0^\pi d\theta \int_{\pi/2}^{3\pi/2} M_s^2 \sin\theta P(\theta,\varphi) d\varphi. \tag{4}$$

Then the error rate is obtained by $1 - P_{\text{switch}}$.

The Fokker-Planck equation (3) can be reduced to the standard eigenvalue problem by the method of separation of variables, which has been extensively studied involves the problems about relaxation time of a nanomagnet or the switching speed of thermally assisted spin torque-induced switching for $\Delta V/k_B T \gg 1$ [21-27], where $\Delta V$ is the energy barrier height between the bistable states. In this paper, the problem we care about is the probability density distribution evolution with respect to time. By moving and merging terms, (3) can be made of the form of an unsteady convection-diffusion equation:

$$\frac{\partial P}{\partial t} = a \frac{\partial^2 P}{\partial \theta^2} + b \frac{\partial P}{\partial \theta} + c \frac{\partial^2 P}{\partial \varphi^2} + d \frac{\partial P}{\partial \varphi} + eP, \tag{5}$$

where $a$, $b$, $c$, $d$, $e$ are the coefficients of each term. The partial differential equation with the form of (5) can be solved numerically by the finite difference method directly. Using fourth order central difference scheme $\frac{1}{h}(-\frac{1}{12}P_{i+2} + \frac{2}{3}P_{i+1} - \frac{2}{3}P_{i-1} + \frac{1}{12}P_{i-2})$ for the first derivative with respect to $\theta$ and $\varphi$, and fourth order central difference scheme $\frac{1}{h^2}(-\frac{1}{12}P_{i+2} + \frac{4}{3}P_{i+1} - \frac{5}{2}P_i + \frac{4}{3}P_{i-1} - \frac{1}{12}P_{i-2})$ for the second derivative with respect to $\theta$ and $\varphi$, where $h$ is the step of spatial coordinates, we constructed a difference scheme of Crank-Nicolson type with fourth order accuracy in space and second order accuracy in time to solve the partial differential equation (5). We used 400 meshes to sample the $\varphi$ interval $(0, 2\pi)$ and 200 meshes to sample the $\theta$ range around the precession angle due to the small precession angle change during the switching process. The time step was set as 0.2 ps. These values were carefully tested to ensure sufficient calculation accuracy.

## 3. Results and discussion

We adopted the values given in [13] for parameters used in the calculations. These values are listed as follows: The damping factor $\alpha$ is 0.01, the saturation magnetization $M_s$ is $1.23 \times 10^3$ emu/cm$^3$, the temperature $T$ is 300 K, the volume $v$ is $1.12 \times 10^{-16}$ cm$^3$, the in-plane coercive field $H_c$ is 25 Oe, the perpendicular anisotropy field $H_{perp}$ is 1400 Oe under zero bias voltage and 600 Oe under the applied voltage pulse. The calculated probability density distribution of the initial state and after a half-period $\tau_{pulse}$ of 0.46 ns under external field $H_{ext}$=700 Oe are shown in figure 2(a) and figure 2(b), respectively. It can be seen that the hot spot of probability density locates around (1.06, 0) at first and then moves to the position around (1.03, π), which indicates a 180° magnetization reversal after the half-period pulse duration.

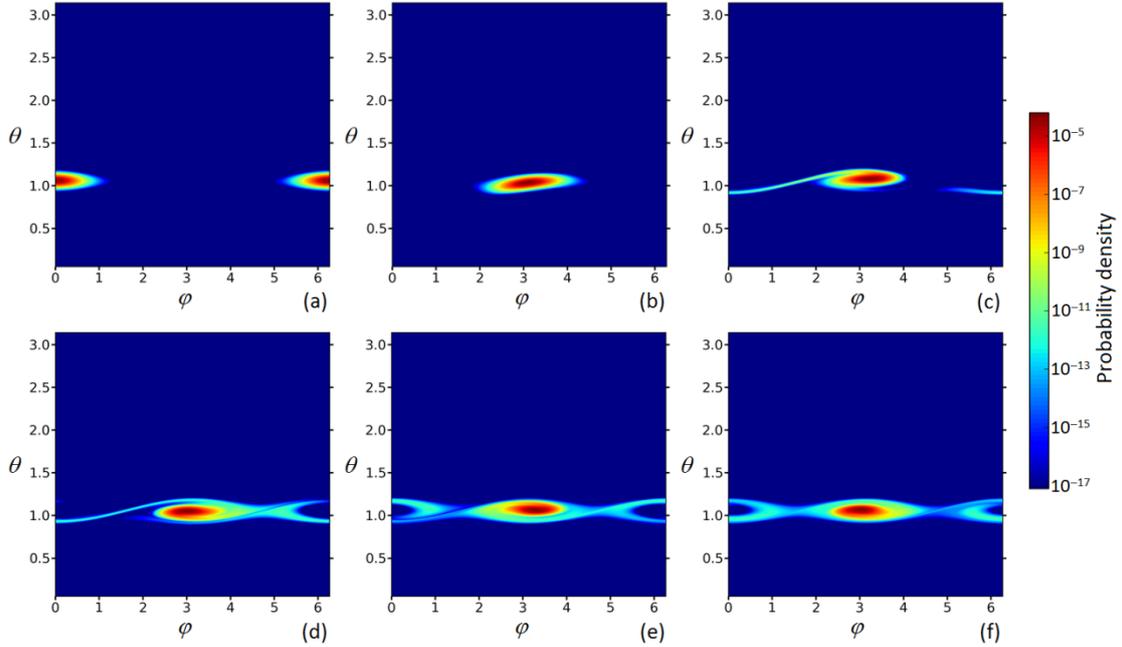

**Figure 2.** The calculated magnetization probability density distribution of (a): the initial state and (b): after half-period pulse duration of 0.46 ns under $H_{ext}$=700 Oe. The hot spot of probability density locates around (1.06, 0) at first and then moves to the position around (1.03, π), which indicates a 180° magnetization reversal after the half-period pulse duration. (c)-(f): Evolution of the probability density distribution with relaxation time after the half-period pulse duration. (c): 1.25 ns; (d): 2.5 ns; (e): 3.75 ns; (f): 5 ns.

From figure 2(b), we can see the probability density distribution incline to the magnetization direction with large $\Delta\theta$ after the precession process. According to the discussion in section 2, this will cause reversed magnetization precession during the relaxation process, and cause switching error. We calculated the evolution of probability density distribution with time during the relaxation process after the half-period $\tau_{pulse}$, as shown in figure 2(c)-(f). The relaxation time is (c) 1.25 ns, (d) 2.5 ns, (e) 3.75 ns and (f) 5 ns, respectively. It can be seen that a portion of probability density flows back to the interval $-\pi/2 \leq \varphi \leq \pi/2$ during the relaxation process, and tends to stabilized gradually with the increase of relaxation time. Using (4), we calculated the error rate as function of the relaxation time, as shown in figure 3(a). It can be seen that the error rate increases up to 1.5 ns and changes little duration under $H_{ext}$ of 700 Oe. (b)-(h): The calculated error rate as function of pulse duration time $\tau_{pulse}$ for varied applied external field $H_{ext}$.

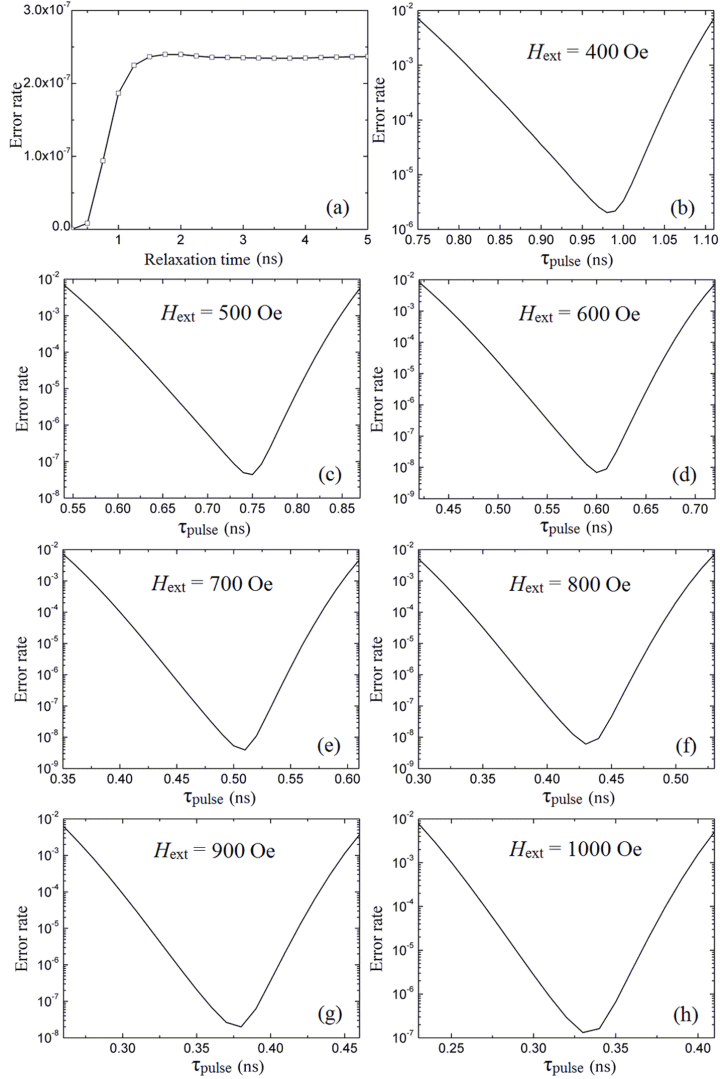

**Figure 3.** (a): The calculated error rate as function of relaxation time after 0.46 ns voltage pulse

afterward. In the calculations below, the switching probability is obtained by integrating probability density distribution $P(\theta,\varphi)$ after sufficient relaxation time to ensure the calculation accuracy.

Figure 3(b)-(h) show the calculated error rate as function of pulse duration time $\tau_{pulse}$ for varied applied external field $H_{ext}$. All $\tau_{pulse}$ ranges are in the vicinity of the half precession period for the applied $H_{ext}$. The precession period is given by $2\pi(1+\alpha^2)/\gamma H$. Ignoring the change of precession angle $\theta$ during magnetization reversal, the estimated half-period time for $H_{ext}$ of 700 Oe is about 0.46 ns. From figure 3(e) for $H_{ext}$=700 Oe, it can be seen that the $\tau_{pulse}$ with the lowest error rate is 0.51 ns. The $\tau_{pulse}$ with the lowest error rate is a little longer than the half-period time of the magnetization precession. This is a general result for varied $H_{ext}$. According to the discussion in section 2, due to the precession axis move during magnetization reversal, the precession angle $\theta$ will decrease first and then increase, so the $\Delta\theta$ has its smallest value at $\varphi=\pi$. Therefore, the $\tau_{pulse}$ with the lowest error rate is nearby the half-period time.

From figure 3, we can see ultra-low error rate less than $4\times10^{-9}$ is achieved for $H_{ext}$=700 Oe. For the $H_{ext}$ higher than 700 Oe, the error rate increases with the increase of $H_{ext}$, and for the $H_{ext}$ lower than 700 Oe, the error rate increases with the decrease of $H_{ext}$. In the rest of this section, we discussed the influence of thermal stability factor, damping factor of free layer and spin transfer torque effect on the switching probability.

*3.1. Influence of thermal stability*

The thermal stability factor $\Delta$ of the free layer is estimated by $\Delta V/k_B T$, where $\Delta V$ is the anisotropy energy barrier height between the bistable states. $\Delta$ determines the initial magnetization probability density distribution before the applied voltage pulse. For the $H_{ext}$ of 700 Oe and $H_c$ of 25 Oe, the estimated $\Delta$ using the parameter values listed in section 3 is about 35. By increasing the $H_c$ from 25 Oe to 36 Oe, a $\Delta$ value of 50 can be achieved. We calculated the error rate as function of $\tau_{pulse}$ for $\Delta$=50 under the $H_{ext}$ of 700 Oe, as shown in figure 4(a). The calculated results show that ultra-low error rate down to $5\times10^{-12}$ is obtained for $\Delta=50$, the switching probability can be greatly improved by enhancing the thermal stability. For large $\Delta$ value, the initial probability density distribution concentrates to the equilibrium position with the lowest potential energy. Therefore, after the precession process, the probability density of magnetization direction with large $\Delta\theta$ decreases compared with the low $\Delta$ value, this leads to the decrease of switching error with the increase of $\Delta$. This result suggests high-reliability write operations can be realized by electric field-induced precessional magnetization switching.

From figure 3 we can see that the error rate increases with the increase of $H_{ext}$ in the high $H_{ext}$ region. The increase of $H_{ext}$ will cause large tilt of magnetization orientation, as shown in figure 1. This decreases the $\Delta V$ and leads to the decrease of thermal stability factor $\Delta$. So the switching probability decreases for $H_{ext}$ higher than 700 Oe.

Then we investigated the temperature dependence of the error rate, which is important for the high-reliability application. In Fokker-Planck equation (3), the thermal agitation term $k=k_B Th/v$ is proportional to the temperature $T$. The change of temperature not only influences the initial probability density distribution, but also influences the distribution during the magnetization precession process. The calculated results for $\Delta=35$ and $H_{ext}=700$ Oe are shown in figure 4(b), it can be seen that the error rate rises by about two orders of magnitude when the temperature rises from 300 K to 400 K. This indicates that the ambient temperature influence on the switching probability can't be ignored.

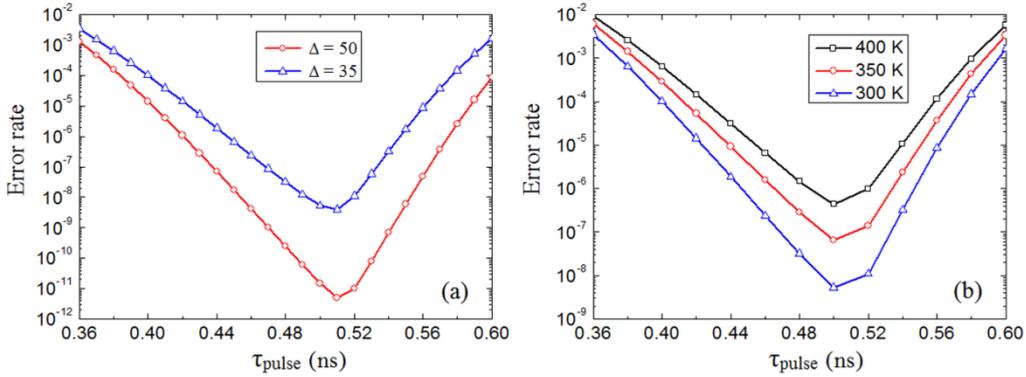

**Figure 4.** (a) The calculated error rate as function of pulse duration time $\tau_{pulse}$ under external field of 700 Oe. Red circle line: thermal stability factor $\Delta$ is 50; Blue triangle line: $\Delta$ is 35. (b) The calculated error rate as function of $\tau_{pulse}$ for external field of 700 Oe and $\Delta$ of 35 for different temperature. Black square line: 400 K; Red circle line: 350 K; Blue triangle line: 300 K.

*3.2. Influence of damping factor α*

During the magnetization precession process, the precession angle $\theta$ is the function of time $t$:

$$\tan\frac{\theta}{2}=e^{-\omega_0 t\alpha/(1+\alpha^2)}\tan\frac{\theta_0}{2}, \tag{6}$$

where $\omega_0$ is the precession angular frequency given by $\gamma H$. Using (6), we can deduce:

$$\tan\frac{\Delta\theta}{2}=\tan(\frac{\theta_0}{2}-\frac{\theta}{2})=(1-e^{-\omega_0 t/(\frac{1}{\alpha}+\alpha)})/(\frac{1}{\tan\frac{\theta_0}{2}}+e^{-\omega_0 t/(\frac{1}{\alpha}+\alpha)}\tan\frac{\theta_0}{2}). \tag{7}$$

From (7) it can be seen that $\Delta\theta$ is an increasing function with respect to $\alpha$ and $\theta_0$ for $\alpha<1$ and $\theta_0<\pi/2$. From (2) we can see $H'$ is also an increasing function with respect to $\theta_0$ and $\Delta\theta$ for $\theta_0<\pi/2$. Therefore, according to (2) and (7), the induced net magnetic field $H'$ after $\tau_{pulse}$ increases with the increase of initial precession angle $\theta_0$.

From figure 3 it can be seen that the error rate increases with the decrease of $H_{ext}$ in the low $H_{ext}$ region. The decrease of $H_{ext}$ will increase the tilt of precession axis from $z$ axis, according to the discussion in section 2, this will increase $\Delta\theta$ after $\tau_{pulse}$. The decrease of $H_{ext}$ will also increase the initial precession angle $\theta_0$, and then increase the $H'$ after $\tau_{pulse}$. Therefore, the decrease of $H_{ext}$ will cause more switching errors. Combining the influence of thermal stability in high $H_{ext}$ region, the error rate will have the lowest value under an appropriate applied external field $H_{ext}$, here is the 700 Oe.

Increasing $H_{ext}$ can decrease $H'$ but it also decrease the thermal stability factor $\Delta$ and the magnetoresistance ratio due to the large magnetization tilt angle. According to (2) and (7), $H'$ decreases with the decrease of damping factor $\alpha$. In Fokker-Planck equation (3), the thermal agitation term $k=k_B Th/v$ is proportional to $h$, which also decreases with the decrease of $\alpha$ for $\alpha<1$. Therefore, to improve the switching probability, we can employ low damping factor $\alpha$ material for the free layer. We calculated the error rate as function of $\tau_{pulse}$ for $\alpha=0.005$ and $0.02$ under the $H_{ext}$ of 700 Oe, as shown in figure 5. It can be seen that ultra-low error rate can be achieved for the low $\alpha$ value, and for $\alpha=0.02$, the error rate rises to the order of $10^{-5}$. It also can be seen that $\tau_{pulse}$ with the lowest error rate is close to the half-period time with the decrease of $\alpha$. These results suggest low damping factor material should be used for the free layer of the electric field-induced precessional magnetization switching.

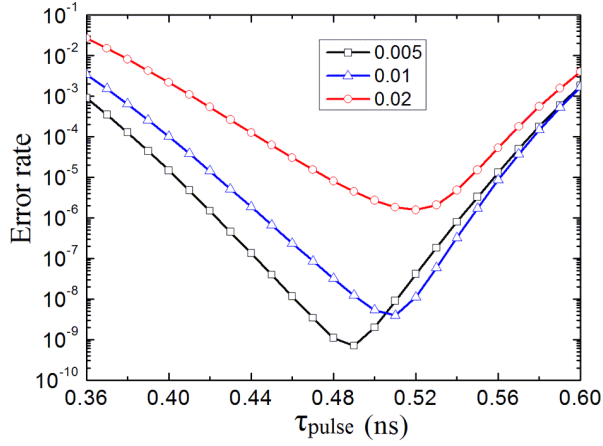

**Figure 5.** The calculated error rate as function of pulse duration time $\tau_{pulse}$ for different value of damping factor $\alpha$ under external field of 700 Oe. Black square line: $\alpha$=0.005; Blue triangle line: $\alpha$=0.01; Red circle line: $\alpha$=0.02.

*3.3. Effect of spin transfer torque*

Finally, we investigated the influence of spin transfer effect on the electric field-induced precessional magnetization switching. In [13], the current density passing through the device when voltage pulse is applied is $2.4\times10^6 A/cm^2$, the torque introduced by the voltage is approximately 68 times that of spin transfer. Though the torque introduced by voltage dominates the precession process, the influence of spin transfer torque (STT) on the switching probability can't be neglected. We added the STT term $-\mathbf{M}\times(\mathbf{M}\times\mathbf{H_s})$ [28] to the Fokker-Planck equation (3), where $\mathbf{H_s}=\left(\frac{J\hbar P}{2etM_s}\right)\mathbf{m_s}$, $J$ ($2.4\times10^6 A/cm^2$) is the current density, $P$ (0.07) is the spin transfer coefficient, $e$ is the electron charge, $t$ ($0.7\times10^{-7}$cm) is the free layer thickness, and $\mathbf{m_s}$ is the unit magnetization vector of the pinned layer, the values in parentheses are from [13].

The calculated results for switching events from parallel (P) to anti-parallel (AP) magnetization configuration and from AP to P magnetization configuration are shown in figure 6(a). It can be seen that the STT effect influences the switching probability remarkably. To explain this, we analyze the dynamic process of magnetization precession. When an external magnetic field is applied, the magnetization vector **M** tilts from *x−y* plane to *z* axis, as shown in figure 1. The spin torque term $-\mathbf{M}\times(\mathbf{M}\times\mathbf{H_s})$ will produce a net torque in the *z* direction. For P to AP, when $-\pi/2<\varphi<\pi/2$ the *z* component of the torque is negative and will decrease the precession angle $\theta$, when $\pi/2<\varphi<3\pi/2$ the $z$ component is positive and will increase $\theta$. If the $z$ component of spin torque is not too large, it will decrease the $\Delta\theta$ when $\varphi<\pi$ and increase $\Delta\theta$ when $\varphi>\pi$, as shown in figure 7(a). This tends to decrease the error rate when $\tau_{pulse}$ is shorter than the half-period, and increase the error rate when $\tau_{pulse}$ is longer than the half-period, as shown in figure 6(a). For AP to P, the situation is just the opposite, the $z$ component of spin torque will increase the $\Delta\theta$ when $\varphi<\pi$ and decrease $\Delta\theta$ when $\varphi>\pi$, as shown in figure 7(c). Therefore, the error rate tends to increase when $\tau_{pulse}$ is shorter than the half-period and decrease when $\tau_{pulse}$ is longer than the half-period. We also calculated the error rate for $J=2.4\times10^5$ A/cm$^2$, as shown in figure 6(b). It can be seen that the influence of STT effect on the switching probability can be ignored when the current density decreases by an order of magnitude [15].

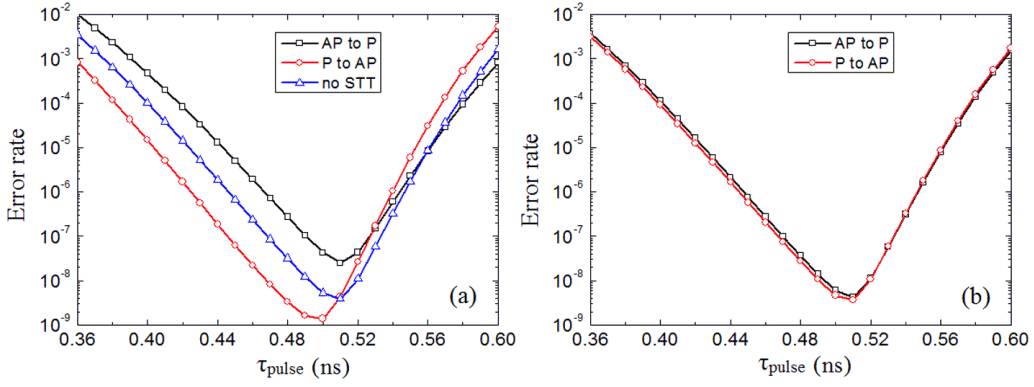

**Figure 6.** The calculated error rate as function of pulse duration time $\tau_{pulse}$ taking into account the spin transfer torque (STT) effect for the applied current density of (a) $2.4\times10^6 A/cm^2$ and (b) $2.4\times10^5 A/cm^2$, respectively. The external field $H_{ext}$ is 700 Oe. Black square line: switching event from anti-parallel (AP) to parallel (P) magnetization configuration; Red circle line: switching event from P to AP; Blue triangle line: no STT effect is taken into account.

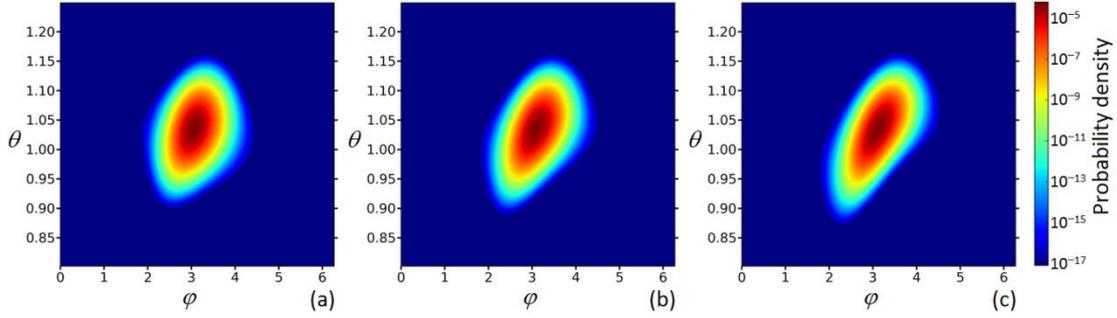

**Figure 7.** The calculated magnetization probability density distribution after 0.46 ns $\tau_{pulse}$ under the external field $H_{ext}$ of 700 Oe. (a): the switching event from parallel (P) to anti-parallel (AP); (b): no STT effect is taken into account; (c): the switching event from AP to P.

## 4. Conclusions

In summary, we investigated the switching probability of electric field-induced precessional magnetization switching with in-plane easy axis by solving the Fokker-Planck equation numerically with finite difference method. The switching probability is determined by the net magnetic field $H'$ induced by the deviation of precession angle $\theta$ from its equilibrium position after precession process. The error rate has the lowest value under an appropriate applied external field $H_{ext}$ for the voltage pulse duration $\tau_{pulse}$ a little longer than the half precession period. The calculated results show that ultra-low error rate down to the order of $10^{-12}$ can be achieved for thermal stability factor $\Delta = 50$ and low damping factor material should be used for free layer to improve the switching probability. For parallel (anti-parallel) magnetization to anti-parallel (parallel) magnetization switching process, the spin transfer torque tends to decrease (increase) the error rate when the $\tau_{pulse}$ is shorter than the half precession period, and increase (decrease) the error rate when $\tau_{pulse}$ is longer than the half-period due to the $z$ direction component of the torque. These results exhibit potential of the electric field-induced precessional magnetization switching for ultra-low power, high speed magnetic random access memory (MRAM) application. We hope this study is helpful to design the electric field-induced precessional magnetization switching device.

**Acknowledgments**

This work was supported by Tsinghua University Initiative Scientific Research Program (No. 20101081760) and Major State Basic Research Development Program 973 (No. 20101972110).